# Multi-Domain Negative Capacitance Effects in Metal-Ferroelectric-Insulator-Semiconductor (Metal) Stacks: A Phase-field Simulation Based Study

Atanu K Saha, and Sumeet K Gupta

*School of Electrical and Computer Engineering, Purdue University, West Lafayette, IN, 47906, USA*

In this work, we analyze the ferroelectric (FE) domain-wall (DW) induced negative capacitance (NC) effect in Metal-FE-Insulator-Metal (MFIM) and Metal-FE-Insulator-Semiconductor (MFIS) stacks. Our analysis is based on 2D phase field simulations, in which we self-consistently solve time-dependent Ginzburg Landau (TDGL) equation, Poisson's equation and semiconductor charge equations. Considering $Hf_{0.5}Zr_{0.5}O_2$ (HZO) as the FE material, we study 180° FE domain formation in MFIM and MFIS stacks and their voltage-dependent DW motion. Our analysis signifies that, when FE is in multi-domain (MD) state with soft-DW, the stored energy in the DW leads to non-hysteretic NC effect in FE, which provides an enhanced charge response in the MFIM stack, compared to Metal-Insulator-Metal. According to our analysis, the DW-induced NC effect yields *local* negative permittivity in FE in the domain and DW regions, which leads to an *average* negative effective permittivity in FE. Furthermore, we show that the NC trajectory of FE is dependent on its thickness, the gradient energy coefficient and the in-plane permittivity of the underline DE material but not on the DE thickness. Similar to MFIM, MFIS also exhibits an enhancement in the overall charge response and the capacitance compared to MOS capacitor. At the same time, the MD state of FE induces non-homogenous potential profile across the underlying DE and semiconductor layer. In the low voltage regime, such non-homogenous surface potential leads to the co-existence of electron and hole in an undoped semiconductor, while at higher voltages, the carrier concentration in the semiconductor becomes electron dominated. In addition, we show that with FE being in the 180° MD state, the *minimum* potential at FE-DE interface and hence, the *minimum* surface potential in the semiconductor, does not exceed the applied voltage (in-spite of the local differential amplification and charge enhancement).

The negative capacitance (NC) effect in ferroelectric (FE) materials has attracted an immense attention because of its potential to overcome the fundamental limits in field-effect transistor (FET) operation[1]. In conventional Metal-Oxide-Semiconductor (MOS)-FET, only a fraction of the applied gate voltage ($V_{APP}$) appears as semiconductor surface potential ($\Psi$) due to a voltage drop across the positive gate dielectric capacitance ($d\Psi/dV_{APP}$<1). Therefore, the attainable subthreshold swing (SS) in MOSFET is always higher than the 60 mV/decade at room temperature (T=300K)[1]. However, it has been proposed that an FE layer at the gate stack of FE-FET (or NC-FET) can act as a negative capacitor and thus, can amplify the internal potential so that $d\Psi/dV_{APP}$>1 and SS < 60 mV/decade at room temperature[1].

According to Landau's free energy equation[1], FE polarization ($P$) vs electric-field ($E$) characteristics exhibit an unstable negative slope. According to ref. 1, such an unstable (negative $dP/dE$) region in FE can be stabilized in a heterogeneous system (i.e. FE-DE stack) so that a homogeneously suppressed polarization ($P$=0) can be obtained by suppressing the depolarization energy. However, under certain conditions it may be more natural to form multiple domains with positive and negative $P$ separated by domain-walls (DWs) to suppress the depolarization energy of the system[2]. Recently, DW motion-based $P$-switching in multi-domain (MD) FE has been identified as a possible mechanism for obtaining static NC in FE[3-5]. Such DW-induced NC effect has been theoretically predicted in ref. 6-7 showing that the soft-DW displacement can lead to an effective negative permittivity of FE in presence of the interfacial dead layer. Further, a similar effect has been analyzed through phase-field simulations predicting a hysteresis-free NC path in FE by considering a moving DW in a FE capacitor[8] and DE-FE-DE superlattice[9]. Additionally, an analytical model for DW-induced NC has been proposed for DE-FE-DE superlattice in ref. 9 suggesting that the NC path is dependent on the DE thickness ($T_{DE}$), which contrasts with the analysis presented in ref.10. However, our phase field simulations show that the DW motion-based NC path in FE is independent of $T_{DE}$, but depends on the in-plane permittivity of the DE layer, which is in agreement with ref. 10. To identify such interdependency of FE NC behavior on the properties of the constituent FE and DE layers in such heterostructures, we extensively analyze DW-induced NC effect in MFIM based on phase field simulations (beyond what has been explored so far) and establish its dependence on FE thickness, gradient energy coefficient, and DE permittivity and thickness. Furthermore, we, for the first time, develop a self-consistent 2D phase-field simulation framework for Metal-Ferroelectric-Insulator-Semiconductor (MFIS) stack. Utilizing our framework, we investigate DW induced NC effect in the MFIS stack, its effect on the semiconductor potential and its dependency on key material/device parameters.

In our phase-field simulation framework, (Fig. 1(a)) we solve the 2D time ($t$) -dependent Ginzburg-Landau (TDGL) equation[9], Poisson's equation and semiconductor charge equations, yielding self-consistent solutions for polarization ($P(x, z, t)$), potential ($\Phi(x, z, t)$) and charge ($\rho(x, z, t)$), where $z$ and $x$ are along the thickness and length of the stack, respectively. For the FE material, we consider $Hf_{0.5}Zr_{0.5}O_2$ (HZO) and the corresponding Landau's free energy coefficients ($\alpha$, $\beta$ and $\gamma$) are extracted from measured $P$-$V$ characteristics[11]. For the gradient coefficient ($g$) of HZO, a range of values are considered as the actual value is still unknown. We assume the spontaneous $P$ direction in FE is along the thickness of the film ($z$-axis), which is parallel to the c-axis of the orthorhombic crystal phase[12-13]. For DE, we consider $SiO_2$, $Al_2O_3$ and $HfO_2$, and for the semiconductor, we consider silicon (Si). The simulation parameters are listed in Fig. 1(b) and this framework is utilized for subsequent analysis of MFIM and MFIS stack (Fig. 1(c)).

Let us start by considering an MFIM stack with an applied voltage ($V_{APP}$=0). It is well known that in MFIM stack, spontaneous polarization ($P$) appears at the FE-DE interface, which leads to a voltage drop across the DE. As a result, an E-field appears in FE opposite to the $P$ direction (called depolarization field, $E_{FE,Z}$), which leads to an increase in the depolarization energy density, $f_{DEP}$ (= $-E_{FE,Z}\times P$). However, $f_{DEP}$ can be suppressed with the formation of periodic 180° domains of alternating $P$-directions (P↑ and P↓)[3-7] as shown in Fig. 2(a) for $T_{FE}$=5nm, $T_{DE}$=2nm ($Al_2O_3$), $g$=1x10$^{-9}$m$^3$V/C. In this multi-domain (MD) state, the magnitude of the local $E_{FE,Z}$ (at a particular point in the FE) is greatly reduced due to stray fields (in-plane E-field, $E_{FE,X}$) between P↑ and P↓ domains, as shown in Fig. 2(b). While this decrease in local $E_{FE,Z}$ is larger near the domain walls (DWs) compared to inside of the domains, the suppression of average $E_{FE,Z}$ is significant across the entire length of the stack (along the *x*-direction). The resultant decrease in $f_{DEP}$, however, comes at the cost of DW energy density ($f_{DW}$), which is comprised of (a) the electrostatic energy density ($f_{ELEC,X}=\epsilon_{FE,X}E_{FE,X}^2$) due to stray fields, where $\epsilon_{FE,X}$ is the in-plane background permittivity of FE and (b) gradient energy density ($f_{GRAD,X}=g\times(dP/dx)^2$) due to the spatial variation in $P$ along the *x*-axis. Subsequently, we will refer to the sum of $f_{GRAD,X}$ and $f_{ELEC,X}$ over the FE region as the DW energy ($F_{DW}=\iint f_{DW}\,dxdz$, where, $f_{DW}=f_{GRAD,X}+f_{ELEC,X}$). Note that the magnitude of $P$ inside of a domain also varies along the z-axis exhibiting a minima at the DE interface and gradually increasing in the bulk FE (away from the DE interface). This induces a bound charge density $\rho_b=-dP(z)/dz$ and further suppresses the $E_{FE,Z}$ (and hence, $f_{DEP}$) inside of the domain. However, this additional suppression of $E_{FE,X}$ occurs at the cost of an increase in $f_{GRAD,Z}$ ($=g\times(dP/dz)^2$). Our simulations show that $f_{GRAD,Z}$ occurs in FE both in the MD (co-existing P↑ and P↓) and poled (either P↑ or P↓) states. In the MD state (achieved by suppressing $f_{DEP}$ at the cost of $f_{DW}$ and $f_{GRAD,Z}$ while minimizing the overall system energy), the intricate interactions of these energy components with each other (and the free energy, $f_{FREE}$) play a key role in determining the NC response of FE and its dependence on the device/material parameters, as discussed subsequently.

Let us first describe the implication of FE thickness ($T_{FE}$) on the formation of MD state. The $P$ configuration of FE in MFIM stack for different $T_{FE}$, shown in Fig. 2(c), suggests that the number of domains (and DWs) increases (within a certain length) with the decrease in $T_{FE}$. As $T_{FE}$ decreases, $f_{GRAD,Z}$ increases as a similar $P$ variation along *z*-axis (i.e. similar $P$ maxima in the bulk and minima in the interface) occurs within a lower $T_{FE}$. One of the possible ways to reduce $f_{GRAD,Z}$ could be decreasing $P$ variation by increasing $P$ magnitude in the interface, but this would increase $f_{DEP}$. On the other hand, when the number of DW increases in FE, the domain width is reduced, which leads to higher penetration of domain wall into the domains. This reduces the $P$ magnitude in the bulk FE and hence, an increase in $f_{GRAD,Z}$ due to lower $T_{FE}$ can be mitigated. In this case, suppression of $f_{DEP}$ becomes more significant inside of a domain (as $P$ decreases in magnitude) and also on an average (as the number of DWs increases). At the same time, with decreasing $T_{FE}$, as the number of DWs increases, the nature of DW changes from hard to soft type (Fig. 2(c)). The term *hard*-DW implies that the spatial variation in $P$ within the DW is abrupt ($dP/dx$ is high). Thus, the DWs and domains are physically separable entities. In contrast, in a *soft*-DW, the $P$ distribution is more gradual ($dP/dx$ is low) and the effects of DW ($f_{GRAD,X}$) diffuses along the length-scale of a domain. However, if $T_{FE}$ is scaled below a critical value, a single domain (SD) state with homogenous $P$=0 stabilizes (Fig. 2(c): $T_{FE}$=2nm), where the suppression of $f_{DEP}$ occurs at the cost of $f_{FREE}$ rather than $f_{DW}$. For suppressing $f_{DEP}$, if $f_{DW}$ is higher than $f_{FREE}$ then the SD state is preferred over the MD state. Similar to the effect of $T_{FE}$, the gradient coefficient ($g$) also determines the number of domains and transition from MD to SD states. As $f_{GRAD,X}$ is one of the components of $f_{DW}$, a decrease in $g$ leads to lower DW energy cost and, thus, the formation of larger number of domains (Fig. 3(d)). Also, the critical $T_{FE}$ (for MD→SD transition) decreases with a decrease in $g$ (Fig. 2(e)) as $f_{DW}$ decreases and therefore, needs a lower $T_{FE}$ to go beyond $f_{FREE}$. Note that if $g$ is very small (~0.1x10$^{-9}$ m$^3$V/C), the critical $T_{FE}$ can potentially become so small (~0.25nm) that the SD state may not be physically realizable (see Fig. 2(e)). Similar to $T_{FE}$, the nature of DW changes from hard to soft type as $g$ increases. This is because, for higher $g$, $dP/dx$ decreases (to compensate for the $f_{GRAD,X} = g\times(dP/dx)^2$) and thus the $P$-distribution becomes more gradual and diffuses within the domain. The nature of DW plays an important role in E-field driven DW motion. To displace the hard-DW, the applied E-field needs to be higher than a critical value (coercive field of DW motion, $|E_{C,DW}|>0$) and therefore, DW motion is hysteretic (due to positive (negative) $E_{C,DW}$ for forward (reverse) DW motion)[14]. In contrast, $|E_{C,DW}|$ is infinitesimally small (~0) for soft-DW[14] and hence, non-hysteretic DW motion is possible. As in this work, our focus is on analyzing the non-hysteretic NC effect, therefore, we restrict our discussion only for soft-DW motion based $P$-switching.

Let us begin by discussing $P$-switching in MFIM stack with soft-DW ($T_{FE}$=5nm, $T_{DE}$=4nm ($Al_2O_3$), $g$=1x10$^{-9}$m$^3$V/C). The simulated charge density ($Q$) vs applied voltage ($V_{APP}$) characteristics is shown in Fig. 3(a). Here, $Q=\left[\int_0^l\{\epsilon_{DE}E_{DE,Z}(x)\}dx\right]/l$ (average charge density), $E_{DE,Z}$ is the $z$-component of E-field at Metal-DE interface and $l$ is the length of the stack. For $|V_{APP}|<2V$, a continuous $Q$-$V_{APP}$ path exists when the FE is in MD state and the $P$-switching takes place through DW motion (see Fig. 3(b)). If $|V_{APP}|$ is increased above ~2V, MD state (P↑↓) switches to the poled state (either P↑ or P↓). Now, with decreasing $|V_{APP}|$, MD state forms from the poled state at a lower $|V_{APP}|$ (~0.9V) and that induces a hysteresis in the $Q$-$V_{APP}$ characteristics. Therefore, for non-hysteretic operation, the MD state needs to be retained by limiting the $V_{APP}$. Interestingly, in the MD state, $Q$ is higher in MFIM stack compared to the MIM (Metal-Insulator-Metal) at the same $V_{APP}$ (Fig. 3(a)). That implies, the effective capacitance of the MFIM stack is higher than MIM. In a static scenario, such a phenomena is only possible if the FE layer acts as an effective negative capacitor ($C_{FE}<0$). The extracted $Q$-$V_{FE,EFF}$ ($V_{FE,EFF}=V_{APP}-QT_{DE}/\epsilon_{DE}$) characteristics (in Fig. 3(a)) shows that the $C_{FE}=dQ/dV_{FE,EFF}$ is indeed negative while FE is in MD state and that implies the effective permittivity of the FE layer, $\epsilon_{FE,EFF}$ ($=T_{FE}\times C_{FE}$) is negative.

The DW-motion induced negative effective permittivity can be described as follows. When $V_{APP}$=0V, the P↓ and P↑ domains in FE are equal in size and the *local* $E_{FE,Z}$ (depolarizing field) is directed opposite to the *local P* (*i.e.* P↓ domains exhibit E↑ and P↑ domains exhibit E↓). Note that $f_{GRAD,X}$ is non-zero inside of the domain (due to DW diffusion in soft-DW) and that causes the $P$ to decrease in magnitude (discussed earlier). Now, with the increase in $V_{APP}$, P↓ domains grow and P↑ domains shrink in size, due to positive stiffness of DW motion[7]. As the DW moves away from P↓ domain and towards the P↑ domain, $f_{GRAD,X}$ in P↓ domain decreases and in P↑ domain increases. Due to this as well as because of positive $V_{APP}$, the magnitude of local $P$ in P↓ domain increases and in P↑ domain decreases. Consequently, our simulation shows that the depolarizing field ($E_{FE,Z}$) in P↓ domain increases and in P↑ domain decreases in magnitude. This implies $f_{DEP}$ increases (decreases) in P↓ (P↑) domain. The increase in $f_{DEP}$ in P↓ domains is possible as it is accompanied by a decrease in $f_{GRAD,X}$. Note that, such $V_{APP}$-induced increase/decrease in $P$, is not directly driven by E-field in the FE; rather, the depolarizing E-field appears depending on the change in $P$ induced by DW motion. As the *oppositely directed* local E-field in FE increases (decreases) with the increase (decrease) in local $P$ in both P↓ and P↑ domains, the *local* permittivity of the domains ($\epsilon_{FE,LOCAL}(x)$) become negative. At the same time, in the DW, the asymmetry in $P$ distribution (due to unequal P↑ and P↓ domain sizes and $P$ magnitudes) causes $F_{DW}$ (comprised of $f_{GRAD,X}$ and $f_{ELEC,X}$) to decrease compared to the symmetric $P$ distribution (at $V_{APP}$=0)[2]. Such decrease in $F_{DW}$ allows a further increase in average-$E_{FE,Z}$ (an increase in depolarization energy) in the DW, while the average-$P$ (directed opposite to $E_{FE,Z}$) in the DW increases (due to unequal $P$ magnitudes in P↑ and P↓ domain). As a consequence, the permittivity of the DW region also becomes negative. These *local* negative permittivity of the domain and DW regions give rise to a negative *average* permittivity in the FE layer (which was earlier referred to as effective permittivity of FE) i.e. $\epsilon_{FE,EFF}<0$ .

As we have identified that the $F_{DW}$ plays a crucial role in providing $\epsilon_{FE,EFF}<0$, therefore, it is easy to understand how the NC behavior is dependent on the $f_{GRAD,X}$ ($=g\times(dP/dx)^2$) and $f_{ELEC,X}$ ($=\epsilon_{FE,X}E_{FE,X}^2$) in the FE. To investigate such dependency, the NC path in the $Q$-$E_{FE,EFF}$ (where, $E_{FE,EFF}=V_{FE,EFF}/T_{FE}$) responses of MFIM stack for different $g$ are shown in Fig. 3(c), which clearly exhibit an increase in the NC effect (increase in $1/|\epsilon_{FE,EFF}|=|dE_{FE,EFF}/dQ|$) with an increase in $g$. As the $f_{GRAD,X}$ increases with the increase in $g$, a higher energy reduction (or gain) can be achieved by displacing the DW, which further provides a higher increase (decrease) in $f_{DEP}$ in P↓ (P↑) domains, leading larger NC effect. Similarly, $dP/dx$ increases as the number of domains and the DWs increase with the decrease in $T_{FE}$ (discussed before). Therefore, $f_{GRAD,X}$ increases and provides an increased NC effect with decreasing $T_{FE}$ (Fig. 3(d)). However, the soft-DW induced NC path does not depend of $T_{DE}$ (Fig. 3(e)). This because, in the MD state, the average depolarization field (which is zero at $V_{APP}$ =0) as well as $f_{GRAD,X}$ and $f_{ELEC,X}$ is independent of $T_{DE}$. Interestingly, the MD-NC path does depend on the DE permittivity ($\epsilon_{DE}$) as shown in Fig. 3(f). This is because the in-plane E-field, $E_{FE,X}$ in the DW needs to satisfy the in-plane boundary condition at the FE-DE interface, which is $E_{FE,X}=E_{DE,X}$ where $E_{DE,X}$ and $E_{FE,X}$ are the in-plane E-field in DE and FE, respectively. As the $E_{DE,X}$ increases with the decrease in $\epsilon_{DE}$ (considering similar $P$ difference between two consecutive domains), therefore, $E_{FE,X}$ also increases in FE, which further increases the $f_{ELEC,X}$ ($=\epsilon_{FE,X}E_{FE,X}^2$) stored in the DW. Therefore, the $F_{DW}$ increases and hence, NC effect increases with the decrease in $\epsilon_{DE}$ as shown in Fig. 3(f). From this analysis, we can summarize that, (i) an FE material with higher $g$, (ii) $T_{FE}$ scaling and/or (iii) using low $\epsilon_{DE}$ DE materials are key device design knobs to enhance DW-induced NC effect (to increase $1/|\epsilon_{FE,EFF}|$). Note that in all of the cases discussed above, the MD NC path does not coincide with the Landau path (Fig. 3(c-f)) and the MD NC effect is less compared to the NC effect that corresponds to Landau path. Now, as the MD NC path is dependent on $T_{FE}$, $g$ and $\epsilon_{DE}$, therefore, the charge enhancement characteristics also depend on them. The charge response in MFIM ($Q_{MFIM}$) and in MIM ($Q_{MIM}$) can be written as $Q_{MFIM} = Q_{MIM}\times(1-\epsilon_{DE}T_{FE}/(T_{DE}|\epsilon_{FE,EFF}|))^{-1}$. Therefore, the charge enhancement increases with the increase in $g$ (as $1/|\epsilon_{FE,EFF}|$ increases), decreases with the increase in $T_{DE}$ and shows mild dependency with the increase in $T_{FE}$ (as an increase $T_{FE}$ leads to decrease in $1/|\epsilon_{FE,EFF}|$) and $\epsilon_{DE}$ (as an increase $\epsilon_{DE}$ leads to decrease in $1/|\epsilon_{FE,EFF}|$) due to counteracting factors.

So far, we discussed how the DW-induced NC effect in FE can enhance the overall charge response of MFIM. Next, we turn our attention to the MFIS stack and to compare the results with conventional MOS capacitor, we also simulate MIIS (Metal-$HfO_2$-$SiO_2$-Si) and MIS (Metal-$SiO_2$-Si) stack. The $Q$-$V_{APP}$ and $C$-$V_{APP}$ (capacitance, $C=dQ/dV_{APP}$) responses are shown in Fig. 4(a-b), which illustrate an enhanced charge and capacitance response of MFIS compared to the MIIS and MIS stack. We attribute this to the effective negative $\epsilon_{FE,EFF}$ of FE that we discussed earlier. Now, to analyze the effects of $T_{FE}$, $Q$-$V_{APP}$ characteristics for different $T_{FE}$ is shown in Fig. 4(c) showing minor enhancement in charge response with the increase in $T_{FE}$. To understand this, a relation can be derived between charge response in MFIS ($Q_{MFIS}$) and in MIS ($Q_{MIS}$) when $\epsilon_{FE,EFF}$ is negative as follows: $Q_{MFIS} = Q_{MIS}\times(1-C_{MIS}T_{FE}/|\epsilon_{FE,EFF}|)^{-1}$. Here, $C_{MIS}$ is the capacitance per unit area of MIS stack. Recall that the NC effect decreases ($1/|\epsilon_{FE,EFF}|$ decreases) with the increase in $T_{FE}$ (discussed for MFIM). However, the increase in $T_{FE}$ dominates over decrease in $1/|\epsilon_{FE,EFF}|$ in the expression of $Q_{MFIS}$ above. Consequently, the charge responses show a mild boost (1.01x) with the increase in $T_{FE}$ (due to two counteracting factors). Similarly, to analyze the effect of $f_{GRAD,X}$, we simulate MFIS stack for different values of $g$. The $Q$-$V_{APP}$ characteristics (Fig. 4(d)) show that the MFIS charge response enhances with the increase in $g$ and are higher than the corresponding MIIS and MIS stack. This is because the NC effect enhances ($1/|\epsilon_{FE,EFF}|$ increases) with the increase in $g$, as we discussed earlier in the context of MFIM.

The overall enhancement in charge/capacitance response of MFIS stack (compared to MIS and MIIS) can be easily understood from the effective negative $\epsilon_{FE,EFF}$ of FE. However, for FEFET operation, it is also important to analyze the semiconductor surface potential ($\Psi$) in MFIS, especially, when FE is in MD state (Fig. 5(a)). In fact, $\Psi$ in MFIS is non-homogeneous as shown in Fig. 5(b) at $V_{APP}$=0V. To understand this, let us consider the potential at the FE-DE interface, $V_{INT}$. Note that in the MD state, E-field in FE, $E_{FE,Z}$ ($\approx(V_{APP}-V_{INT})/T_{FE}$) is directed opposite to the local $P$ and exhibits a non-homogeneous profile along the x-direction due to periodic P↑ and P↓ domains. Therefore, $V_{INT}$ becomes non-homogenous and exhibits a maxima (max-$V_{INT}$) and minima (min-$V_{INT}$) corresponding to the center of P↓ and P↑ domains, respectively. This non-homogeneity in $V_{INT}$ induces a spatially varying $\Psi$ (Fig. 5(b)) which, in turn exhibits a maxima (max-$\Psi$) and minima (min-$\Psi$). This further leads to local accumulation and co-existence of electrons or holes in the undoped Si layer (Fig. 5(c)). Note, such a spatially varying charge profile has been experimentally shown in ref. 15 for FE-semiconductor interface, when FE is in MD state. Now, with the increase in $V_{APP}$ (~1.2V), P↓ domains grow and P↑ domains shrink in size leading to an overall increase in average $P$ (Fig. 5(d)). Simultaneously, min/max-$V_{INT}$ increases (Fig. 5(i)) and at the same time exhibit a differential amplification ($dV_{INT}/dV_{APP}$>1) as shown in Fig. 5(h). Here the local differential amplification in min/max-$V_{INT}$ can be attributed to the *local* negative permittivity of FE in the P↓ and P↑ domains (discussed for MFIM). Now, as $V_{INT}$ increases, $\Psi$ everywhere at the Si interface increases and becomes positive (but still remains non-homogeneous, see Fig. 5(e)). Therefore, electron density ($n$) dominates over hole density ($p$) locally and globally (Fig. 5(f)). Note that the increase in $n$ causes the non-homogeneity in $\Psi$ to decrease (Fig. 5(f)) compared to $V_{APP}$=0V (Fig. 5(c)). The $\Psi$ for MFIS, MIIS and MIS stacks for $V_{APP}$=0V and 1.2V are shown in Fig. 5(b) and 5(e). At $V_{APP}$=0V, the max(min)-Ψ in MFIS is higher(lower) than the MIIS and MIS stacks. At $V_{APP}$=1.2V, the max-$\Psi$ in the MFIS is higher than the MIIS and MIS and the min-$\Psi$ in MFIS is higher than MIIS but lower than the MIS. This can be understood from the following discussion. As in the MD state, $E_{FE,Z}$ (≈($V_{APP}$-$V_{INT}$)/$T_{FE}$) is directed opposite to the local $P$, therefore, the max-$V_{INT}$ is larger than $V_{APP}$ (for P↓ domains with E↑ i.e. $E_{FE,Z} < 0$) and the min-$V_{INT}$ remains less than $V_{APP}$ (for P↑ with E↓ i.e. $E_{FE,Z} > 0$). This holds true when the FE is in 180$^o$ MD state and an only exception to this (where min-$V_{INT}$>$V_{APP}$ can occur) is for a very small voltage window just before the MD state switches to poled state. Hence, as long as FE remains in the 180$^o$ MD state (i.e. does not switch to the poled state), the min(max)-$V_{INT}$ is always lower(higher) than $V_{APP}$ in MFIS (see Fig. 5(g)). Note that, this statement is also true for MFIM. Now, in the MIS stack, DE layer potential is directly driven by $V_{APP}$ and hence $V_{INT}$=$V_{APP}$. Therefore, min-$V_{INT}$ of MFIS is always less than $V_{INT}$ (=$V_{APP}$) of MIS. In addition, $d\Psi/dV_{INT}$ is <1 and equal for both MFIS and MIS due to the same positive capacitance of the DE layer. As a consequence, the min-$\Psi$ of MFIS is inevitably lower than the $\Psi$ of MIS, when the FE is in 180$^o$ MD state. However, in MIIS, the $V_{INT}$ ($HfO_2$-$SiO_2$ interface potential) is not directly driven by $V_{APP}$ and due to the positive capacitance of the $HfO_2$ layer, $dV_{INT}/dV_{APP}$<1 and $V_{INT}$<$V_{APP}$ (Fig. 5(g-h)). Now, considering the differential amplification of min-$V_{INT}$ in MFIS ($d$(*min-*$V_{INT}$)/$dV_{APP}$>1) as shown in Fig. 5(h), the min-$V_{INT}$ of MFIS becomes higher than the $V_{INT}$ of MIIS beyond a certain $V_{APP}$ (Fig. 5(h)). As a result, min-$\Psi$ of MFIS becomes higher than the $\Psi$ of MIIS at $V_{APP}$>1V) as shown in Fig. 5(h). Briefly, in MFIS, the min-$\Psi$ can exceed the $\Psi$ in MIIS but remains lower than the MIS, while the max-$\Psi$ is always higher than the $\Psi$ in MIIS and MIS.

Now, let us make a rough assumption that the channel current in FEFET will be mostly dependent on the min-$\Psi$ as that is the highest potential barrier seen by the source electrons. Then, based on the above discussion, we can expect that the OFF current (at $V_{APP}$=0V) of FEFET will be significantly less compared to the MIS/MIIS-FET, and the ON current ($V_{APP}$~1.2V) will be higher than the MIIS-FET but comparable to MIS-FET. As the Ψ is highly non-homogeneous in MFIS stack in the low voltage regime, calculation of SS of FEFETs needs further exploration by considering source/drain regions along with the DW-induced non-homogenous semiconductor potential and solving the transport equations to obtain the impact of MD FE on the FEFET characteristics.

In summary, by performing phase field simulation, we show that the energy stored in FE DW can be harnessed to enhance the capacitance of the MFIM and MFIS stack, where the soft-DW displacement leads to a static and hysteresis-free negative capacitance in MD FE. Our analysis indicate that the effective negative permittivity of the FE layer is dependent on the FE thickness, gradient energy coefficient, in-plane permittivity of the DE and is independent of DE thickness. Further, the DW-induced NC can lead to an enhanced charge/capacitance response in MFIS stack compared to MIS/MIIS stack. However, such a charge/capacitance enhancement in MFIS does not guarantee an enhanced local $\Psi$ in Si compared to MIS. In fact, $\Psi$ becomes spatially varying due to the MD nature of FE and the variation is higher at low applied voltages. In addition, we discuss that the minimum $\Psi$ in MFIS can exceed the $\Psi$ in MIIS but remains smaller than the MIS. Nevertheless, considering the local differential amplification of $V_{INT}$ (*i.e.* $d(min\text{-}V_{INT})/dV_{APP}>1$), the on/off current ratio of FEFET can potentially exceed the MIS/MIIS-FET. Since the non-homogeneity in $\Psi$ is absent in conventional MOS capacitor (and MOSFET), therefore, as future work, it will be important to investigate the impact of such potential profile on the low voltage conduction of FEFETs.

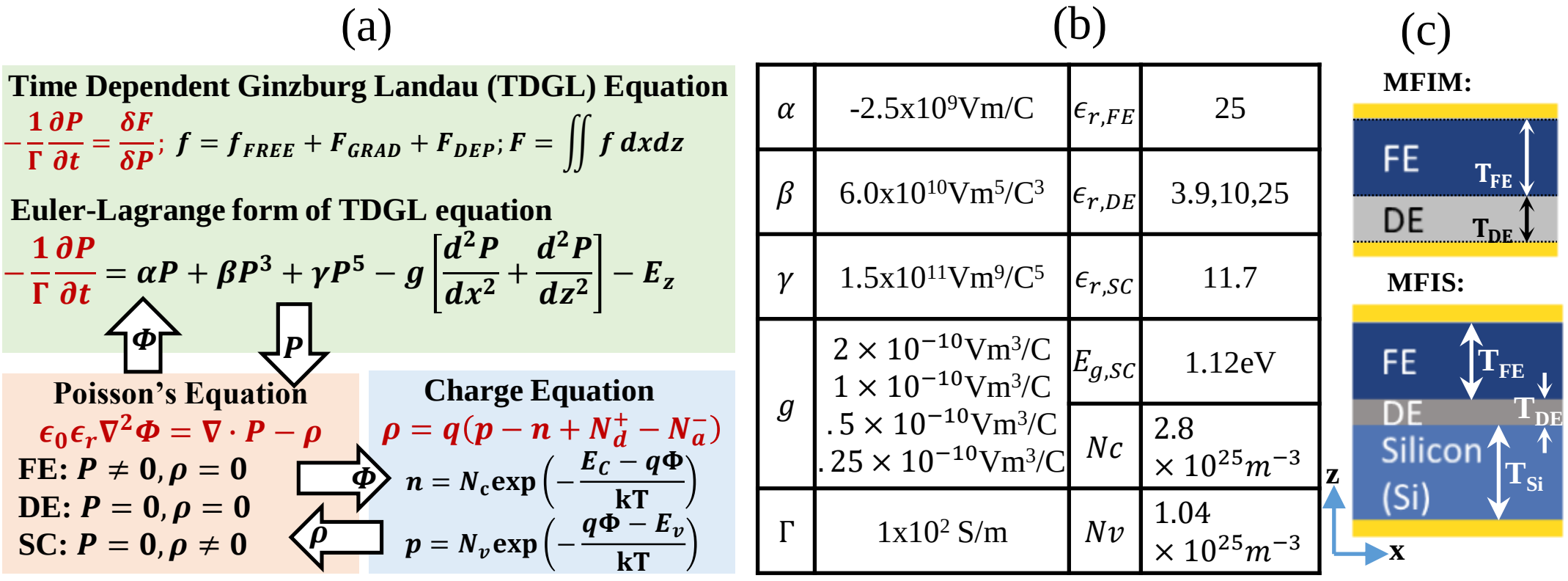

| | | | |
|---|---|---|---|
| $\alpha$ | -2.5x10$^9$Vm/C | $\epsilon_{r,FE}$ | 25 |
| $\beta$ | 6.0x10$^{10}$Vm$^5$/C$^3$ | $\epsilon_{r,DE}$ | 3.9,10,25 |
| $\gamma$ | 1.5x10$^{11}$Vm$^9$/C$^5$ | $\epsilon_{r,SC}$ | 11.7 |
| $g$ | $2\times10^{-10}$Vm$^3$/C<br>$1\times10^{-10}$Vm$^3$/C<br>$.5\times10^{-10}$Vm$^3$/C<br>$.25\times10^{-10}$Vm$^3$/C | $E_{g,SC}$ | 1.12eV |
| | | $Nc$ | $2.8\times10^{25}m^{-3}$ |
| $\Gamma$ | 1x10$^2$ S/m | $Nv$ | $1.04\times10^{25}m^{-3}$ |

Fig. 1: (a) Simulation framework for ferroelectric based devices that self-consistently solves the time-dependent Ginzburg Landau equation (TDGL) with device electrostatics (Poisson's equation) and semiconductor charge equation. (b) Simulation parameters and (c) MFIM and MFIS configuration.

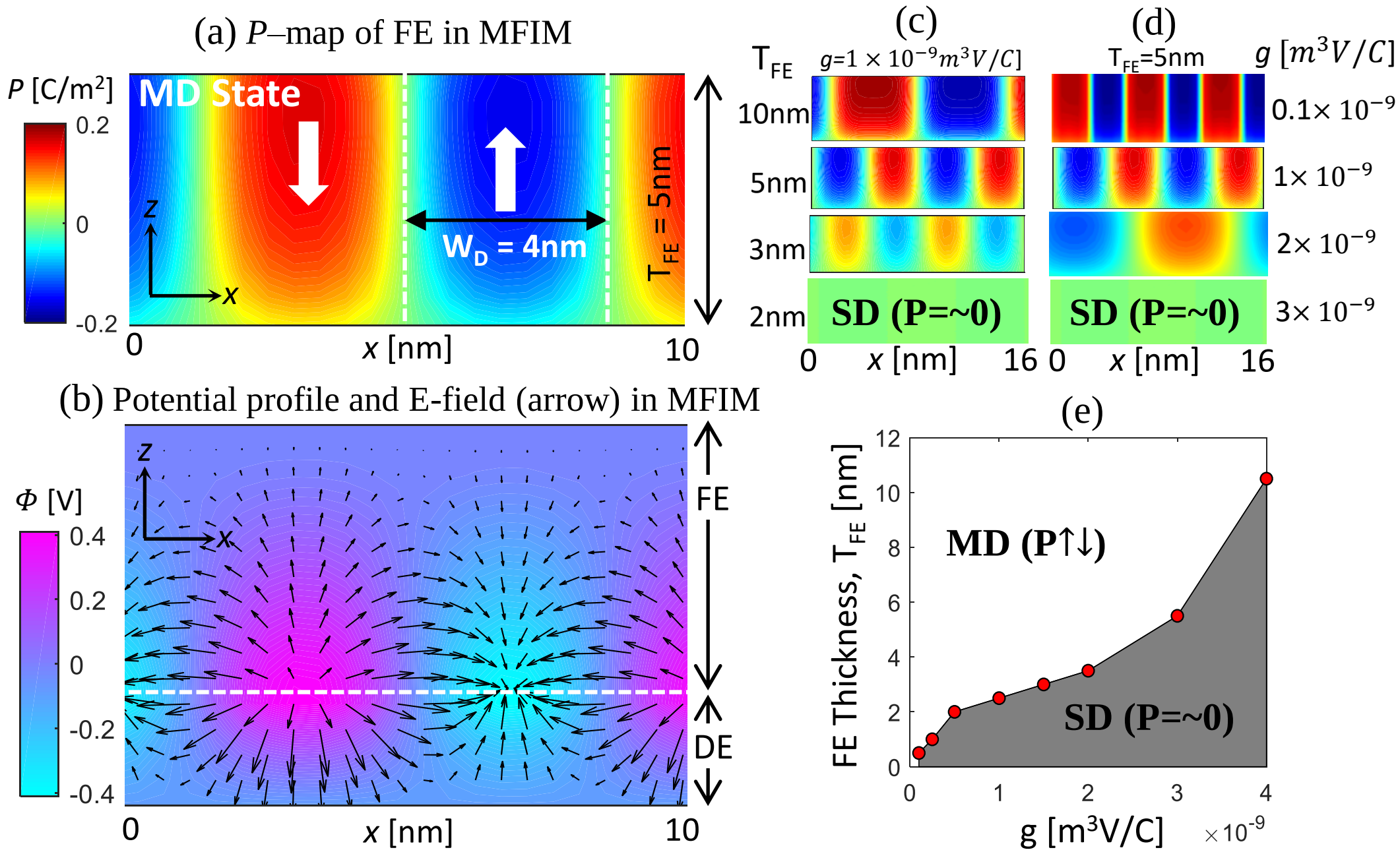

Fig. 2: (a) $P$–map ($P(x,z)$) of FE in MFIM and (b) potential profile ($\Phi(x,z)$) at $V_{APP}$ = 0V showing the formation of 180$^o$ domain structure. Here, $T_{FE}$ = 5nm, $T_{DE}$ = 2nm, $\epsilon_{r,DE}$=10 ($Al_2O_3$). $P$–map of FE in MFIM for (c) different $T_{FE}$ and (d) different $g$ at $V_{APP}$ = 0V. (e) $g$ vs critical $T_{FE}$ below which SD (P~0) state is preferred over MD (P↑↓) state considering $\epsilon_{r,DE}$=10. In the $P$-maps, the $P$-direction is ↓ (↑) in red: +$P$ (blue: -$P$) regions.

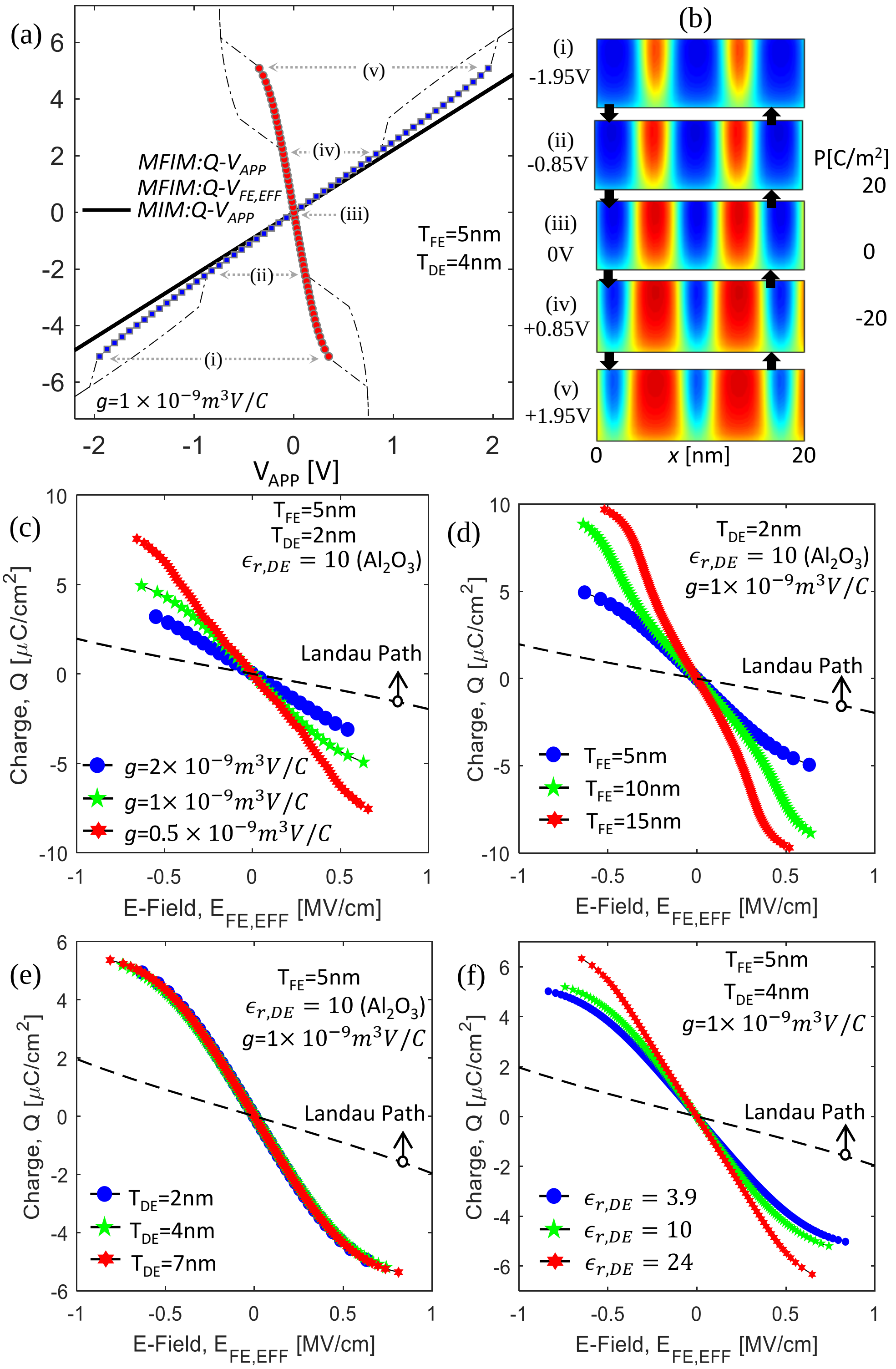

Fig. 3: (a) $Q$-$V_{APP}$ characteristics of MFIM stack when FE is in MD state (blue) showing enhanced charge response of MFIM stack compared to MIM (DE only: black-solid). The black-dashed line represents the poled condition (if $V_{APP}$>2V). Extracted $Q$-$V_{FE,EFF}$ response (red-circle). (b) $P$–map of FE at different $V_{APP}$ (as marked in Fig.3 (a)). $Q$-$E_{FE,EFF}$ characteristics of FE in MFIM stack considering (c) different $g$, (d) different $T_{FE}$, (e) different $T_{DE}$ and (f) different $\epsilon_{DE}$. Black-dashed line in (c-f) is the landau path.

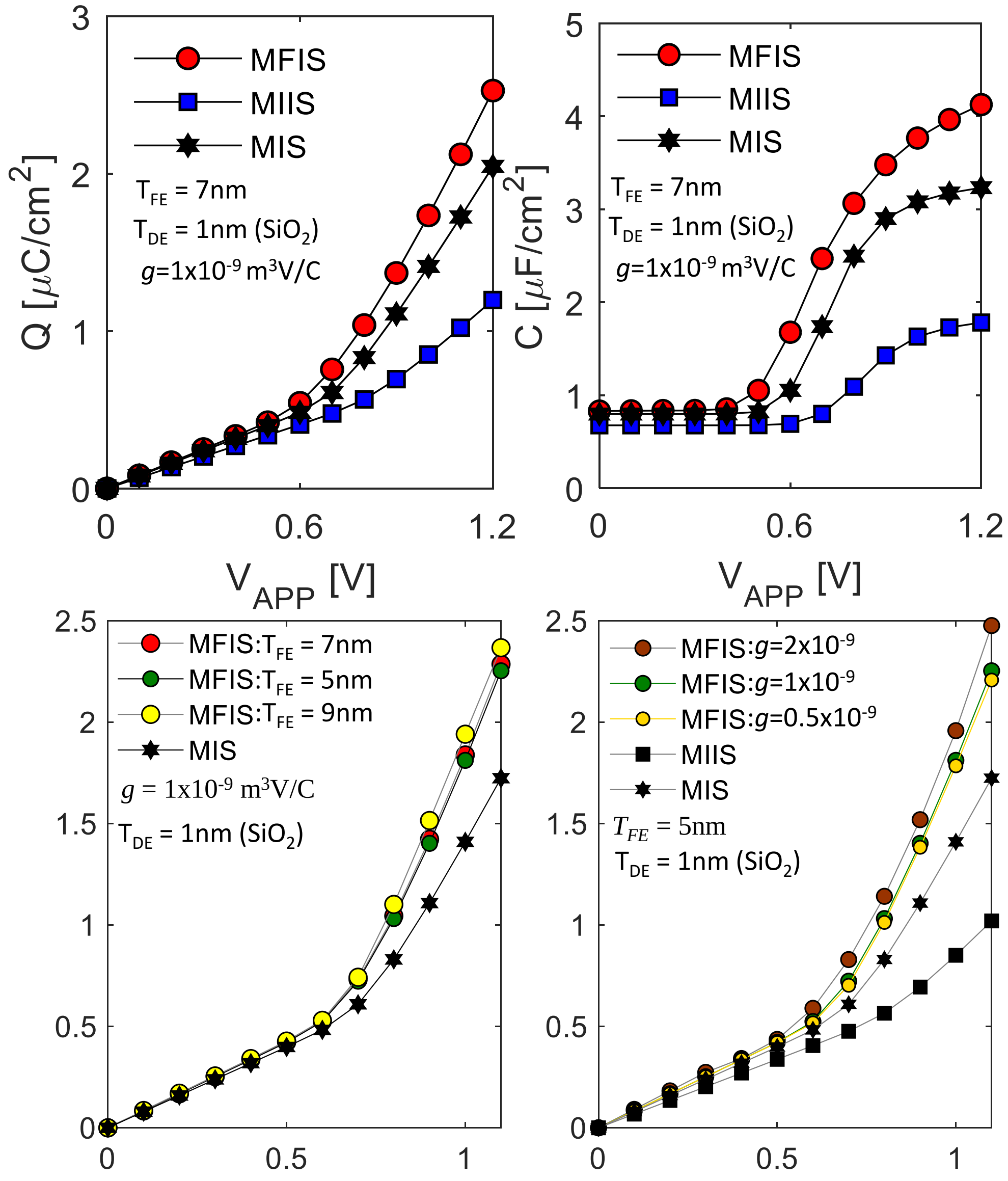

Fig. 4: (a) $Q$-$V_{APP}$ and (b) $C$-$V_{APP}$ of MFIS (HZO(5nm)-$SiO_2$(1nm)-Si(10nm)), MIS ($SiO_2$(1nm)-Si(10nm)), MIIS ($HfO_2$(5nm)-$SiO_2$(1nm)-Si(10nm)) stacks. Here, $g$=1x10$^{-9}$ m$^3$V/C. (c) $Q$-$V_{APP}$ of MFIS stack for different $T_{FE}$ and comparison with MIS stack. Here, $g$=1x10$^{-9}$ m$^3$V/C. (d) $Q$-$V_{APP}$ of MFIS stack for different $g$ and comparison with MIIS, MIS stack. Here, $T_{FE}$=5nm.

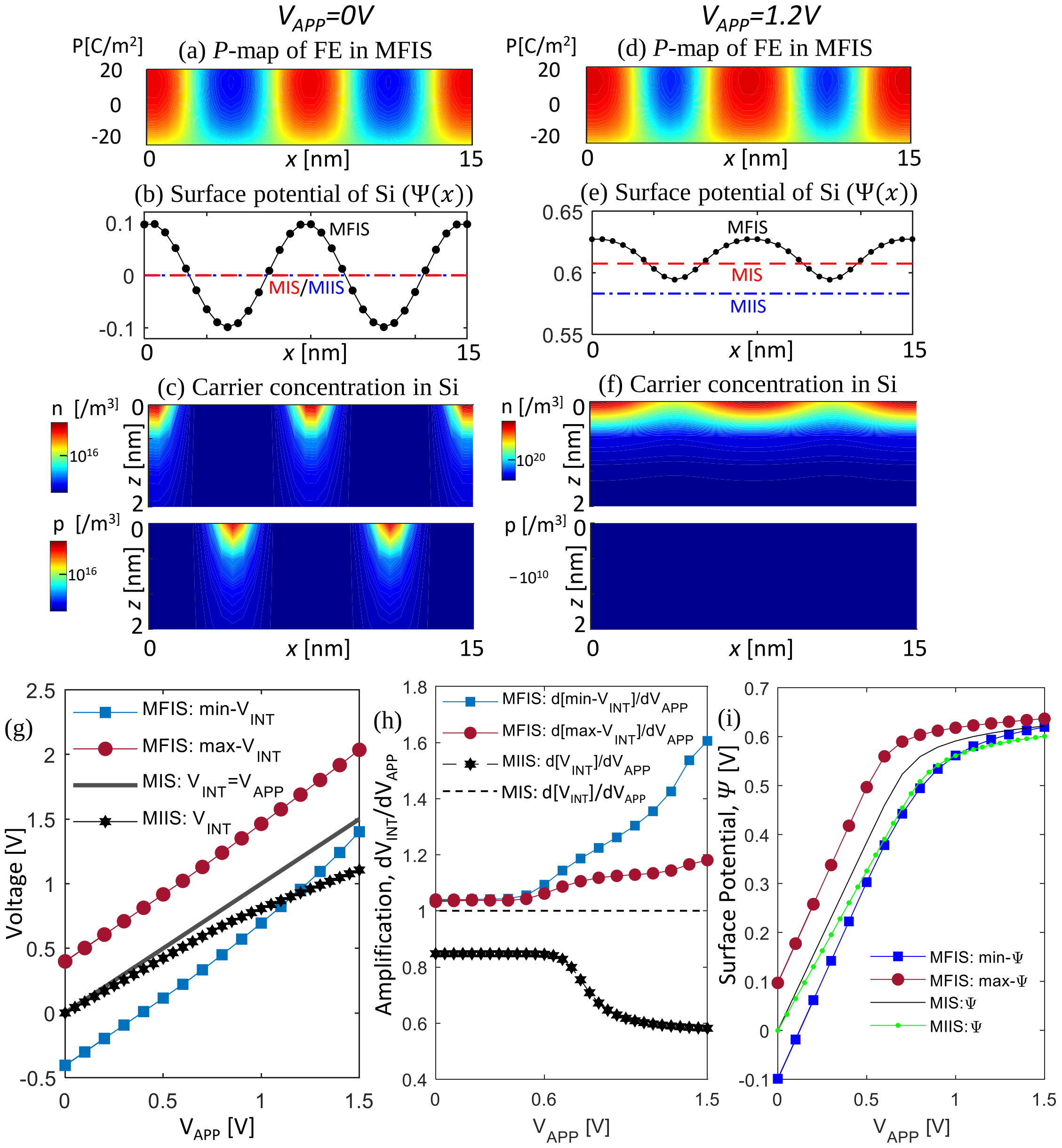

Fig. 5: (a) $P$-map in FE layer, (b) surface potential (Ψ) of silicon, and (c) electron concentration (n) and hole concentration (p) in Si layer at $V_{APP}$ = 0V. (d) $P$-map (e) Ψ and (f) $n$ and $p$ in Si layer at $V_{APP}$ = 1.2V. (g) $V_{INT}$ of MFIS, MIS and MIIS stack. (h) differential amplification of $V_{INT}$ in MFIS, MIS and MIIS stack. (i) Ψ in MFIS, MIS and MIIS stack. Here, $T_{FE}$=5nm, $T_{DE}$=1nm ($SiO_2$) and $g$=1x10$^{-9}$ m$^3$V/C.